\documentclass[review]{elsarticle}
\pdfoutput=1
\usepackage{lineno,hyperref}
\usepackage{multirow}
\usepackage{amssymb}
\usepackage{graphicx}
\usepackage{array}
\usepackage[figuresright]{rotating}
\journal{Journal of \LaTeX\ Templates}

\begin{document}

\begin{frontmatter}

\title{Adversarial Attacks and Defenses in Speaker Recognition Systems: A Survey\tnoteref{mytitlenote}}

%% or include affiliations in footnotes:
\author[mymainaddress]{Jiahe Lan}

\author[mymainaddress]{Rui Zhang}

\author[mymainaddress,mysecondaryaddress]{Zheng Yan\corref{mycorrespondingauthor}}
\cortext[mycorrespondingauthor]{Corresponding author}
\ead{zheng.yan@aalto.fi}

\author[mymainaddress]{Jie Wang}

\author[mythirdaddress]{Yu Chen}

\author[mymainaddress]{Ronghui Hou}

\address[mymainaddress]{State Key Laboratory on Integrated Services Networks, School of Cyber Engineering, Xidian University, China}
\address[mysecondaryaddress]{Department of Communications and Networking, Aalto University, Finland}
\address[mythirdaddress]{School of Information Systems and Technology, Lucas College and Graduate School of Business, San Jose State University, USA}

\begin{abstract}
Speaker recognition has become very popular in many application scenarios, such as smart homes and smart assistants, due to ease of use for remote control and economic-friendly features. The rapid development of SRSs is inseparable from the advancement of machine learning, especially neural networks. However, previous work has shown that machine learning models are vulnerable to adversarial attacks in the image domain, which inspired researchers to explore adversarial attacks and defenses in Speaker Recognition Systems (SRS). Unfortunately, existing literature lacks a thorough review of this topic. In this paper, we fill this gap by performing a comprehensive survey on adversarial attacks and defenses in SRSs. We first introduce the basics of SRSs and concepts related to adversarial attacks. Then, we propose two sets of criteria to evaluate the performance of attack methods and defense methods in SRSs, respectively. After that, we provide taxonomies of existing attack methods and defense methods, and further review them by employing our proposed criteria. Finally, based on our review, we find some open issues and further specify a number of future directions to motivate the research of SRSs security.
\end{abstract}

\begin{keyword}
\texttt{speaker recognition system, adversarial attacks, adversarial examples.}
\end{keyword}

\end{frontmatter}

%\linenumbers

\section{Introduction}
Biometrics such as fingerprint, face, and voiceprint is widely used for user identification and authentication \cite{rui2018survey, yan2016usable}. Speaker recognition, as a technology that recognizes a speaker’s identity through his/her voiceprint \cite{zheng2017robustness,wang2021attacks}, has attracted special attention from both academia and industry due to its ease of use for remote control and economic-friendly features. The last decade has seen a dramatic improvement in Speaker Recognition Systems (SRSs), which can be divided into Speaker Identification Systems (SISs) and Speaker Verification Systems (SVSs) according to different tasks. The former is responsible for identifying which enrolled speaker utters an input, and the identification result is an enrolled speaker. The latter aims to verify whether an input is uttered by a claimed speaker, and the verification result is yes or no. SRSs have been deployed in both classical and emerging Internet-of-Things (IoT) devices \cite{zhang2022volere}, such as smartphones, laptops, smart speakers, and smart homes.

The rapid development of SRSs is inseparable from the advancement of Neural Networks (NNs), especially Deep Neural Networks (DNNs). While SRSs based on traditional methods, such as i-vector \cite{dehak2009support} and Gaussian Mixture Model (GMM) \cite{rasmussen1999infinite}, have prospered for decades, they are being replaced by NN-based methods due to the strong ability of NNs. However, previous work has demonstrated that NNs are susceptible to adversarial attacks \cite{szegedy2013intriguing}. Adversarial attacks mean that an adversary utilizes adversarial examples, which are generated by adding small perturbations, i.e., adversarial perturbations, into clean samples, to make a machine learning model misbehave. Adversarial attacks were first conducted in the image field. Szegedy et al. \cite{szegedy2013intriguing} successfully fooled an image classification model using adversarial examples. After that, adversarial attacks have gained widespread attention in the image field and many effective attack methods, such as Fast Gradient Sign Method (FGSM) \cite{goodfellow2014explaining} and Basic Iterative Method (BIM) \cite{kurakin2016adversarial}, have been proposed. Defense methods have also been extensively studied, such as feature squeezing \cite{xu2017feature} and adversarial training \cite{goodfellow2014explaining}. Akhtar et al. \cite{akhtar2018threat} and Yuan et al. \cite{yuan2019adversarial} comprehensively reviewed existing adversarial attack methods and defense methods in the image field, respectively.

Inspired by the advancement of adversarial attacks in the image field, an increasing number of researchers pay their attention to adversarial attacks in the audio field. As the most widely used voice processing system, the speech recognition system was successfully deceived by adversarial examples in 2015 \cite{vaidya2015cocaine}. Three years later, Kreuk et al. \cite{kreuk2018fooling} first successfully attacked an SRS through FGSM, which proves the effectiveness of adversarial attacks in SRSs. Since then, adversarial attacks and defenses in SRSs started to draw special attention.

Several researchers have surveyed the security of SRSs from different perspectives. Wu et al. \cite{wu2015spoofing} presented a study on spoofing attacks, including impersonation, relay, speech synthesis, and voice conversion, and countermeasures in SVSs. However, adversarial attacks and defenses were not mentioned in their work. SISs were also missed. Das et al. \cite{das2020attacker} demonstrated the security vulnerabilities of SVSs from the perspective of attackers by considering both spoofing attacks and adversarial attacks. Nevertheless, they ignored SISs and did not consider how to evaluate the performance of different attack methods or defense methods. Abdullah et al. \cite{abdullah2021sok} researched adversarial attack methods on both speaker recognition and speech recognition comprehensively. However, they paid little attention to defense methods. Overall, a comprehensive review on the recent advance of adversarial attacks and defenses in SRSs shall be done.

In this paper, we fill the gap by providing a thorough survey on adversarial attacks and defenses in SRSs. We first illustrate the basics of SRSs. At the same time, we introduce several basic concepts and threat models related to adversarial attacks and defenses. After that, we propose taxonomies of attack methods and defense methods respectively. The former includes optimization-based attacks and signal processing-based attacks, and the latter involves proactive defenses and passive defenses. Based on classification and evaluation criteria, we comprehensively review attack methods and defense methods in SRSs and analyze their pros and cons. Finally, we figure out several unsolved issues and suggest future research directions. Specifically, the contributions of this paper are as follows:

\begin{itemize}
\item This paper proposes two sets of evaluation criteria for adversarial attacks and defenses in SRSs, respectively.
\item This paper provides taxonomies of existing adversarial attacks and defense methods.
\item Based on the taxonomies, this paper comprehensively reviews the existing attacks and defenses in SRSs, and evaluates them by employing our proposed criteria.
\item This paper points out several open issues and suggests future research directions based on systematic reviews.
\end{itemize}

The rest of this paper is organized as follows. Section 2 gives a brief overview of SRSs and introduces the basic concepts related to adversarial attacks and defenses. In Section 3, we propose two sets of criteria for evaluating adversarial attacks and defenses, respectively. Section 4 provides the taxonomy of adversarial attacks in SRSs and conducts a thorough review of existing attacks, followed by the taxonomy of countermeasures and a thorough review on them in Section 5. In Section 6, we highlight open issues and propose future research directions. Finally, we conclude this paper in the last section.

\section{Preliminaries}

In this section, we first introduce the basics of SRSs followed by the definitions of the three important concepts in this paper, i.e., adversarial examples, adversarial perturbations and adversarial attacks. Finally, we list possible threat models in adversarial attacks and defenses.

\subsection{Speaker Recognition Systems Overview}

\begin{figure}[tbh]
	\centering
	\includegraphics[width=\textwidth]{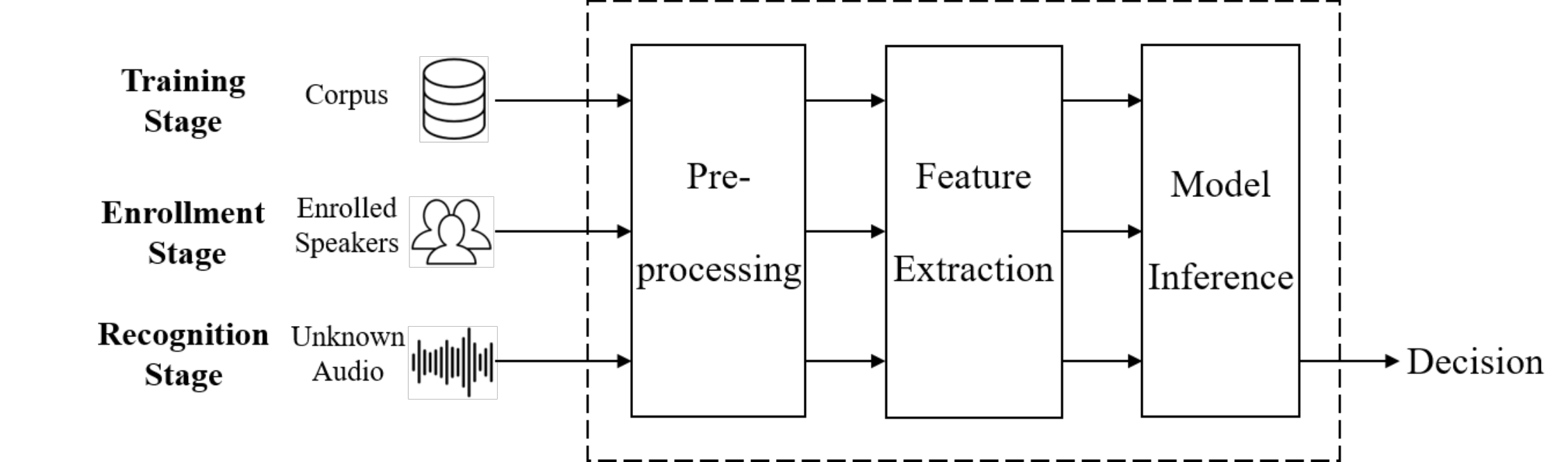}
	\caption{An Overview of A Typical SRS}
	\label{fig:1}
\end{figure}

Fig. 1 shows an overview of a typical SRS. The SRS includes three modules, i.e., preprocessing module, feature extraction module, and model inference module. Meanwhile, the lifecycle of an SRS involves three stages, i.e., training stage, enrollment stage, and recognition stage.

When audio is input, the preprocessing module first filters out background noise and high-frequency signals beyond the frequency range of human voices. Feature extraction algorithms are then used to generate a feature vector that reduces dimensions of the audio by capturing its most important features and characteristics. Various feature extraction algorithms have been proposed, such as Mel-Frequency Cepstral Coefficients (MFCC) \cite{tiwari2010mfcc}, Spectral Subband Centroid (SSC) \cite{paliwal1998spectral}, and Perceptual Linear Predictive (PLP) \cite{hermansky1990perceptual}. Among them, MFCC is the most popular one in SRSs due to its ability to expose important acoustic features, similar to human ears. After that, the feature vector is passed to a model for either training or inferencing.

In the training stage, corpora are used to train the SRS and adjust system parameters to obtain a capable SRS. After that, multiple speakers enroll in the SRS. All enrolled speakers form a speaker group. The SRS calculates and stores a feature vector for every enrolled speaker, which is used in the recognition stage. In the recognition stage, the SRS is responsible for recognizing the identity of unknown input audio.

According to the difference of recognition tasks, SRSs can be divided into two categories: Speaker Verification and Speaker Identification. For an arbitrary input audio $x$, a Speaker Identification System (SIS) determines which enrolled speaker utters $x$. Regarding a Speaker Verification System (SVS), an unknown speaker not only needs to input his/her audio $y$ but also needs to claim his/her identity. The SVS determines whether $y$ is uttered by the claimed speaker, and the verification result is “yes” or “no”.

\subsection{Adversarial Examples Generation}

\emph{Adversarial attack} is a kind of attack method that an adversary generates adversarial examples to make a machine learning model misbehave. An \emph{adversarial example} refers to specifically crafted input designed to look normal to humans but causes misbehaviors of a machine learning model \cite{szegedy2013intriguing}. Given an input $x$
with its corresponding label $y$, and a well-trained machine learning model $f(\cdot)$, an adversarial example $x'$ can be constructed as:

% MathType!MTEF!2!1!+-

\[x' = x + \delta {\rm{ }} \wedge {\rm{ }}f(x',\theta) \ne y{\rm{ }} \wedge {\rm{ }}\left\| \delta  \right\| < \varepsilon \]

Here, $\delta$ is called \emph{adversarial perturbation}, which is the noise that is added to a clean sample to make it an adversarial example \cite{szegedy2013intriguing}. $\theta$ is the parameter of the machine learning model $f(\cdot)$. The hyperparameter $\epsilon$ is used to control the maximum perturbation generated. Suppose $ L(\cdot)$ is the loss function and $y'$ is the target label, the adversarial perturbation $\delta$ can be calculated by
% MathType!MTEF!2!1!+-
% feaahqart1ev3aaatCvAUfeBSjuyZL2yd9gzLbvyNv2CaerbuLwBLn
% hiov2DGi1BTfMBaeXatLxBI9gBaerbd9wDYLwzYbItLDharqqtubsr
% 4rNCHbWexLMBbXgBd9gzLbvyNv2CaeHbl7mZLdGeaGqiVu0Je9sqqr
% pepC0xbbL8F4rqqrFfpeea0xe9Lq-Jc9vqaqpepm0xbba9pwe9Q8fs
% 0-yqaqpepae9pg0FirpepeKkFr0xfr-xfr-xb9adbaqaaeGaciGaai
% aabeqaamaabaabauaakqaabeqaaiGac2gacaGGPbGaaiOBaiaadYea
% caGGOaGaamOzaiaacIcacaWG4bGaey4kaSIaeqiTdqMaaiykaiaacY
% cacaWG5bGaai4jaiaacMcaaeaacaWGZbGaaiOlaiaadshacaGGUaWa
% auWaaeaacqaH0oazaiaawMa7caGLkWoacqGH8aapcqaH1oqzaaaa!570E!
\[\begin{array}{l}
\min L(f(x + \delta,\theta ),y')\\
s.t.\left\| \delta  \right\| < \varepsilon 
\end{array}\]

To solve the above formula, many attack methods have been proposed. We first introduce a classic and well-known method, FGSM \cite{goodfellow2014explaining}. The adversarial perturbation can be calculated by the following formula.
\[\delta  = \eta  sign({\nabla _x}f(x,\theta ))\]

Here, $\eta$ is the magnitude of the perturbation. The adversarial example $x'$ is calculated as: $x'=x+\delta$.
Kurakin et al. \cite{kurakin2016adversarial} proposed BIM which iterates FGSM for multiple rounds. The adversarial example is generated in multiple iterations.

\[x{'_0} = x\]

\[x{'_{n + 1}} = Clip(x{'_n} + \eta sign({\nabla _x}f(x{'_n},\theta )))\]

Here, $Clip(\cdot)$ is a function which limits the change of the generated adversarial example in each iterations.

\subsection{Threat Models}

In this subsection, we introduce threat models of attack methods and defense methods, respectively.
\subsubsection{Threat Models of Attack Methods}

The threat model of an attack method is related to the adversary’s goal, which is also called adversarial specificity, and the adversary’s knowledge about target attack models. Based on the adversary’s goal, adversarial attacks can be categorized into targeted attacks and untargeted attacks. Based on the adversary’s knowledge about target attack models, attacks can be divided into white-box attacks, grey-box attacks, and black-box attacks. The specific definitions are as follows.

\emph{Targeted Attacks}: since an SVS only has two possible decisions (i.e., yes or no), we regard all adversarial attacks on SVSs as targeted attacks. In terms of an SIS, the targeted attack means a speaker is maliciously identified as a specific speaker.

\emph{Non-targeted Attacks}: it means a speaker fails to be identified or is identified as any other speaker in an SIS.

\emph{White-box Attacks}: the adversary has full knowledge about a target model, such as model architecture, parameters, gradients, layer outputs, input and output pairs, etc.

\emph{Grey-box Attacks}: the adversary only knows part of knowledge about a target model.

\emph{Black-box Attacks}: the adversary can only get the input and output pairs of a target model.

\subsubsection{Threat Models of Defense Methods}

The threat model of a defense method is related to the adversary’s knowledge about defense methods. Based on the adversary’s knowledge about defense methods, attacks can be divided into adaptive attacks and non-adaptive attacks. Their specific definitions are as follows.

\emph{Adaptive Attacks}: the adversary has full knowledge about a deployed defense method. Using this knowledge, the adversary can adjust its attack methods to overcome the deployed defense method. But a strong defense method can still counter this type of attack.

\emph{Non-adaptive Attacks}: the adversary does not have any knowledge about a deployed defense method. Resisting non-adaptive attacks is the minimum requirement that a feasible defense method needs to meet.

\section{Evaluation Criteria}

In this section, we propose two sets of evaluation criteria (As shown in Fig. 2). One is used to evaluate attack methods against SRSs, thus analyzing the performance of each attack method. The other is used for evaluating defense methods, thus analyzing the performance of each defense method.

\subsection{Evaluation Criteria for Attack Methods}

In this subsection, we put forward evaluation criteria for attack methods from three aspects: practicability, imperceptibility, and effectiveness.

\subsubsection{Practicability}

Practicability refers to the ability of an attack method to be used in the real world. We introduce the following five metrics to evaluate the practicability of an attack method.

\emph{Transferability}: it is the ability of an adversarial example to continue to make an impact on SRSs other than the one created it. In the real world, SRSs are usually black-box to adversaries. Thus, investigating transferable attack methods is meaningful since they can be used to attack different SRSs. According to the difference between two SRSs before and after transferring, transferability can be classified into cross-feature, cross-dataset, cross-model, etc. Cross-feature, cross-dataset, and cross-model indicate that two SRSs differ from feature extraction technologies, training datasets, and model frameworks, respectively.

\emph{Universality}: it is the ability of an adversarial perturbation to fool a given model on any clean samples with high probability. If an adversary can generate a universal adversarial perturbation, he can obtain adversarial examples by adding the perturbation to any clean samples effortlessly, which helps to achieve real-time attacks.

\emph{Attack Media}: adversarial examples can be fed into an SRS via different medium, each of which introduces different challenges such as background noise and distortion. There are three common attack media including over-line, over-air, and over-telephone-network. Over-line attacks indicate that a Waveform Audio file is directly fed to an SRS. They are the easiest to execute since over-line ensures lossless transmission. Over-air attacks indicate that an audio file played by a speaker is fed to an SRS. Although over-air attacks are more difficult to implement than over-line attacks since the quality of adversarial examples will decrease due to the background noise, attenuation, and multi-path effects during transmission, they are close to the real world. Over-telephone-network attacks are more difficult to implement than over-air attacks since adversarial examples not only pass through the air but also the telephone network. Serious signal processing operations, such as jitter and compression, will further reduce the quality of adversarial examples. To summarize, the overall difficulty of achieving adversarial attacks in the above media is over-line<over-air<over-telephone-network.

\emph{Distance}: the farther the adversarial example travels in loss medium, the worse the audio quality. Therefore, distance is used to measure the farthest distance an adversarial example can spread without losing its ability to attack.

\emph{Commercial SRSs}: it is used to evaluate whether an attack method can successfully attack commercial SRSs, such as Azure Verification API \cite{AVA} and Azure Attestation API \cite{AIA}, which are more complex and have higher security requirements. Once a commercial SRS is attacked successfully, the security and privacy of its users and the reputation of the vendor will be severely damaged.

\subsubsection{Imperceptibility}

Imperceptibility indicates that adversarial perturbations impair utterances very slightly for human perception. Since if ordinary people can distinguish an adversary example and its corresponding clean sample, the adversarial example is likely to be discarded before entering SRSs. We introduce the following four metrics to evaluate the imperceptibility of adversarial perturbations.

\emph{Types of Adversarial Audio}: depending on the attack type and scenario, adversaries can produce different types of adversarial audio. Adversarial audio can be categorized into three classes, including noisy, clean, and inaudible audio. Noisy audio sounds like noise to humans but is considered legitimate audio by SRSs. Clean audio is perturbed at such low intensity that human listeners cannot perceive these perturbations at all, even though there is a hidden command embedded in it. However, SRSs can detect and execute these embedded commands. Inaudible audio is generated by an adversary exploiting the characteristics of human auditory system. On the one hand, the human auditory system can only perceive frequency that ranges from 20Hz to 20kHz. Therefore, audio whose frequency is beyond 20kHz cannot be heard by humans but can be recognized by SRSs. On the other hand, frequency masking which refers to the phenomenon that one faint but audible sound becomes inaudible in the presence of another louder audible sound can also be used to generate inaudible adversarial audio. To summarize, the imperceptibility in the above audio types is noisy audio\textless clean audio\textless inaudible audio.

\emph{Perturbation Norm}: it indicates the restricted $l_p-norm$ of the perturbations to make them imperceptible.$l_2$ and $l_\infty$ are two commonly used metrics. $l_2$  measures the Euclidean distance between the clean sample and the corresponding adversarial example, and $l_\infty$ denotes the maximum change direction between the clean sample and the corresponding adversarial example.

\emph{Human Perception}: it is used to evaluate whether human perception has been considered, in other words, whether user studies (e.g., ABX test) have been implemented in a paper. Human perception is the most direct metric to measure the imperceptibility of adversarial examples.

\emph{Signal-to-noise Ratio(SNR)}: it is defined as the ratio of signal power to the noise power and is often expressed in decibels (dB). The noise is the adversarial perturbations. A larger SNR value indicates a smaller perturbation.

\subsubsection{Effectiveness}

Effectiveness is used to quantitatively measure the performance of an attack method. We introduce the following six metrics to evaluate the effectiveness of attack methods. Generation time and attack success rate are direct metrics to measure the effectiveness of an attack method. The remaining four metrics are indirect metrics since they are used to evaluate the performance of an SRS. The difference between these four metrics before and after the SRS is attacked can be used to measure the effectiveness of an attack method.

\emph{Generation Time}: it refers to the time required to generate an adversarial example. The less the generation time, the more efficient the attack method.

\emph{Attack Success Rate(ASR)}: for targeted attacks, ASR refers to a proportion of adversarial examples that are recognized as the targeted speaker. For non-targeted attacks, ASR indicates a proportion of adversarial examples that are recognized as other speakers rather than the original speaker.

\emph{Recognition Accuracy(RA)}: it is a proportion of utterances correctly recognized by an SRS.

\emph{False-positive Rate (FPR)}: it is a proportion of utterances of non-original speakers that are recognized as the original speaker.

\emph{False-negative Rate (FNR)}: it is a proportion of utterances of original speakers that are recognized as non-original speakers.

\emph{Equal Error Rate (EER)}: it refers to the rate when FPR is equal to FNR. The lower the EER, the greater the performance of an SRS.

\subsection{Evaluation Criteria for Defenses}

From the perspective of practicability and effectiveness, we propose a set of criteria to evaluate the performance of defense methods in SRSs.

\subsubsection{Practicability}

Practicability refers to the ability of a defense method to be used in the real world. We introduce the following three metrics to evaluate the practicability of a defense method.

\emph{Generality}: it is the ability of a defense method to resist different attack methods. In the real world, designers of SRSs usually cannot know which attack methods will be used by adversaries in advance. Therefore, the more attack methods a defense method can resist, the more practical it is. Based on the number of attack methods that can be resisted, we divide defense methods into three levels: high-, medium-, and low-generality. Firstly, highly general defense methods can resist any attack methods. Because attack-agnostic defense methods do not rely on any attack methods, they can be used to resist any attack methods theoretically. Therefore, attack-agnostic defense methods are high-generality. Secondly, in addition to attack-agnostic defense methods, there are some defense methods that rely on attack methods. For example, a defense method needs adversarial examples, which can be generated by FGSM or other attack methods, to adjust parameters. In terms of them, the transferable defense method is medium-generality since it can not only detect adversarial examples generated by the attack method it depends on, but also detect adversarial examples generated by other attack methods. At last, the non-transferable defense method, which can only resist adversarial examples generated by the attack method it depends on, is low-generality.

\emph{Defense Media}: similar to attack media, there are three common defense media including over-line, over-air, and over-telephone-network. In Subsection 3.1.1, we analyze that the overall difficulty of achieving adversarial attacks in the above media is over-line\textless over-air \textless over-telephone-network. The overall difficulty of achieving defenses in the above media is over-line \textgreater over-air \textgreater over-telephone-network.

\emph{Defendable Attacks}: for a defense method, its corresponding defendable attacks refer to the attacks that can be resisted, as proved by experiments. The more attacks that can be resisted, the more effective the defense method is.

\subsubsection{Effectiveness}

Effectiveness is used to quantitatively measure the performance of a defense method. Six metrics are introduced to evaluate the effectiveness of a defense method. Defense time and detection accuracy are direct metrics, while, the same as Subsection 3.1.3, the remaining four metrics, including RA, FPR, FNR, and EER, are indirect metrics. We can evaluate the ability of a defense method through the difference between the four metrics before and after deploying the defense method. We do not repeat the definition of RA, FPR, FNR, and EER in this subsection.

\emph{Defense Time}: it refers to the time required for a defense method to detect or purity an input utterance.

\emph{Detection Accuracy (DA)}: it refers to the ratio of correct discrimination on adversarial examples. The higher the DA is, the more effective the defense method is.

\begin{figure}[htp]
	\centering
	\includegraphics[width=\textwidth]{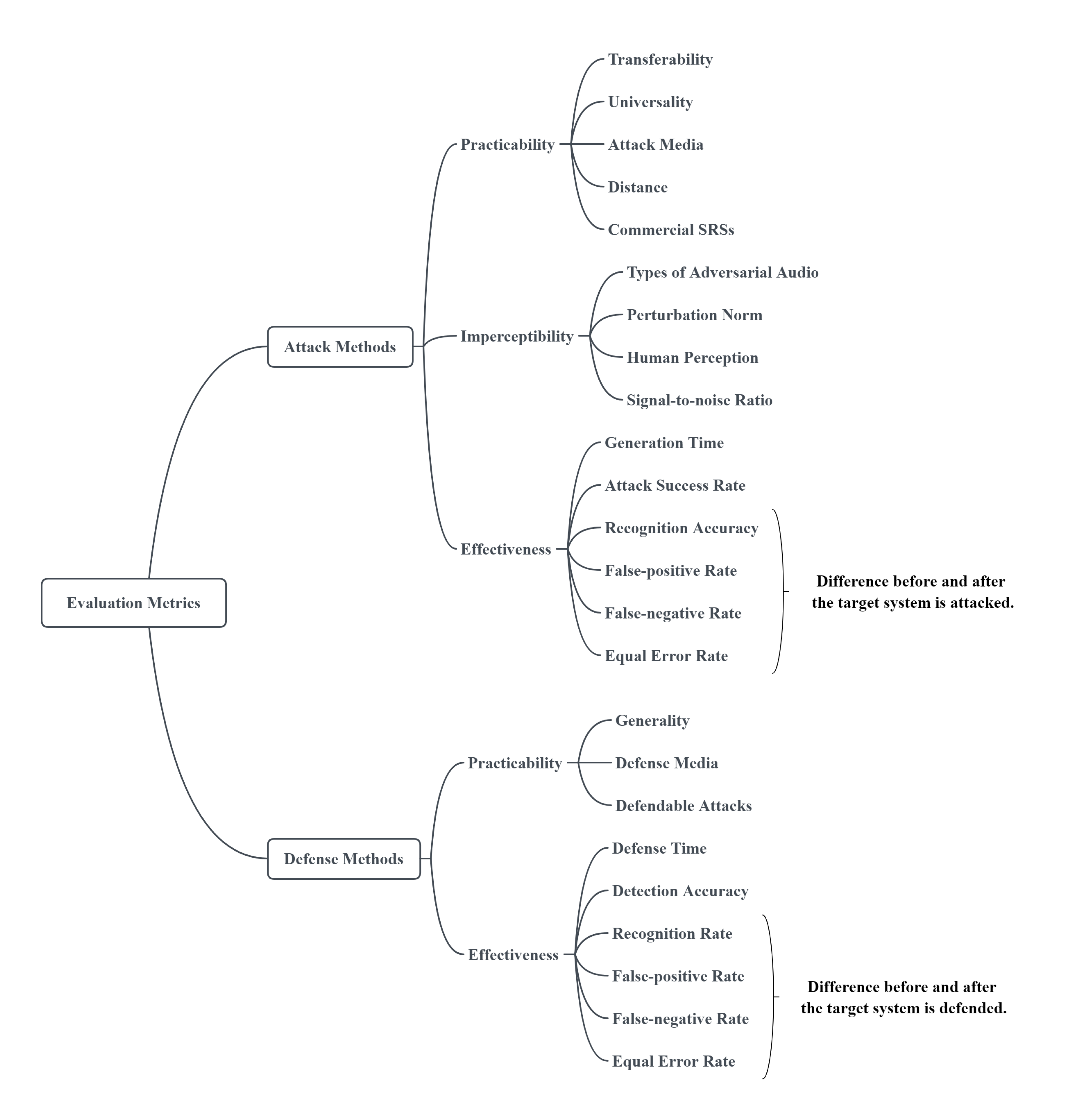}
	\caption{Evaluation Criteria of Attack and Defense Methods}
	\label{fig:1}
\end{figure}

\section{Adversarial Attacks against SRSs}

In this section, we first propose a taxonomy of existing adversarial attack methods against SRSs. Then, we review them based on the taxonomy. Furthermore, we evaluate and compare these attack methods (as shown in Table 1) with our proposed criteria in Subsection 3.1.

\subsection{Taxonomy of Adversarial Attacks against SRSs}

Adversarial attacks against SRSs can be categorized as optimization-based attacks and signal processing-based attacks. The optimization-based attacks generate adversarial examples by solving an optimization problem that is obtained by formalizing the purpose of adversarial attacks. In recent years, several optimization-based adversarial example generation algorithms have been proposed, such as FGSM \cite{goodfellow2014explaining} and BIM \cite{kurakin2016adversarial}. The signal processing-based attacks use signal processing techniques to generate adversarial examples. Although these attacks do not directly target machine learning models embedded in SRSs, they can still force machine learning models to misbehave.

\subsection{Optimization-based Attacks}

Inspired by the success of adversarial attacks in computer vision and speech recognition, Kreuk et al. \cite{kreuk2018fooling} tried to fool a DNN-based SVS by adversarial examples for the first time. Adversarial examples were generated by FGSM, which is $ l_\infty$ perturbation norm. They verified the effectiveness of adversarial attacks in SVSs by a white-box attack. Then they deployed two black-box attacks to explore the transferability of adversarial examples they generated. Experimental results show that the adversarial examples are transferable in both cross-feature and cross-dataset settings. Finally, through human perception experiments, they found that human listeners cannot distinguish between clean samples and adversarial examples, which reflected the imperceptibility of adversarial perturbations generated by FGSM. However, since the purpose of this work was to verify that SVSs are susceptible to adversarial examples, the authors did not make efforts to pursue a better attack effect and did not attack commercial SRSs. In addition, they ignored generation time and over-air attacks, which are more practical than over-line attacks.

Li et al. \cite{li2020adversarial} found that \cite{kreuk2018fooling} has two limitations. On the one hand, \cite{kreuk2018fooling} studied the impact of adversarial attacks on DNN-based SVSs, but ignored GMM-based SVSs. On the other hand, cross-model transferability was not discussed. To overcome the two limitations, they deployed a white-box attack to verify that GMM-based SVSs are also subject to adversarial examples. They then implemented three black-box experiments to study the transferability of adversarial examples generated by FGSM. The results show that adversarial examples are transferable in cross-feature, cross-dataset, and cross-model settings. Finally, they demonstrated the imperceptibility of adversarial examples they generated to human listeners by human perception experiments. Although \cite{li2020adversarial} solved two limitations of \cite{kreuk2018fooling}, the attack performance of \cite{li2020adversarial} was not good enough since it generated adversarial examples by simple FGSM. In addition, they did not discuss generation time and over-air attacks.

The aforementioned work cannot achieve real-time attacks since they generated different adversarial perturbations for different audio, which usually costs a lot of time. Real-time attacks are inseparable from universal adversarial perturbations since universal adversarial perturbations can be added to any samples to generate effective adversarial examples. If an adversary obtains universal adversarial perturbations in advance, it only needs one addition operation to generate an adversarial example, which is time-efficient. Li et al. \cite{li2020universal} tried to generate universal adversarial perturbations for SincNet \cite{ravanelli2018speaker}, a state-of-the-art speaker identification model, with a generative network, which can learn the mapping from a low-dimensional normal distribution to a universal adversarial perturbation subspace. They conducted both untargeted attacks and target attacks against SincNet \cite{ravanelli2018speaker} in grey-box settings. The results show the existence of universal adversarial perturbations. In addition, they studied the imperceptibility of universal adversarial perturbations by SNR. However, they did not deploy over-air attack experiments and did not try to attack commercial SRSs.

Xie et al. \cite{xie2020real} almost simultaneously conducted similar work. While Li et al. \cite{li2020universal} generated universal perturbations with a generative network, Xie et al. \cite{xie2020real} generated universal perturbations with a conventional optimization-based approach. They attacked an SIS in white-box settings successfully. After that, they tried to attack the SIS in the real world, i.e., over-air attacks. However, they failed because of the attenuation and multi-path effects of sound in the propagation process. They then assumed that the adversary has the knowledge of the room’s layout and took Room Impulse Response (RIR) into consideration to enhance the robustness of adversarial examples. The results show that considering RIR can greatly increase the ASR in over-air attacks. However, the high attack success rate came at the cost of low SNR. That means they need to make a trade-off between the attack effect and the imperceptibility of the adversarial perturbations.

Although \cite{li2020universal,xie2020real} generated universal adversarial perturbations successfully, Li et al. \cite{li2020advpulse} found that they only focused on static-speech attack scenarios, but ignored streaming-speech attack scenarios. In static-speech attack scenarios, an adversary should obtain universal adversarial perturbations and clean samples to generate adversarial examples before feeding adversarial examples into an SRS. On the contrary, in streaming-speech scenarios, an SRS takes streaming audio inputs (e.g., live human speech) and an adversary can fool the SRS by playing universal adversarial perturbations through a nearby loudspeaker. There is no doubt that streaming-speech scenarios are closer to the real world than static-speech scenarios. Therefore, Li et al. \cite{li2020advpulse} designed AdvPulse, a method to generate a subsecond-level adversarial perturbation that can be added at any point of a streaming audio input to launch targeted adversarial attacks. In other words, they generated universal adversarial perturbations against SRSs in streaming-speech attack scenarios. In addition, when a loudspeaker and an SRS are less than three meters apart, adversarial examples they generated can successfully deceive the SRS. This work is advanced. Unfortunately, they only considered white-box settings but ignored black-box settings.

Chen et al. \cite{chen2021real} proposed an attack method named Fakebob, which formalizes the generation of adversarial examples into an optimization problem. In this optimization problem, a score threshold and the strength of adversarial perturbations are considered. To solve the optimization problem, they proposed an approach that utilizes a novel algorithm to estimate the threshold, a natural evolution strategy (NES) to estimate gradient, and finally the BIM method is applied to generate adversarial examples. This work is currently the most comprehensive one. It has four main contributions. Firstly, they effectively implemented targeted attacks on SRSs in black-box settings and the ASR can reach 99\%. Secondly, they considered all possible SRSs based on different tasks, including SVSs and SISs. Thirdly, they did many experiments, including over-line and over-air experiments, on both open source systems and commercial systems, which prove the transferability and practicability of Fakebob. Fourthly, they tried four defense methods to defend against Fakebob. Experimental results show that Fakebob still affects the performance of victim systems, which demonstrates the robustness of Fakebob. Meanwhile, they employed human perception experiments to explore the imperceptibility of adversarial perturbations. However, FakeBob took several minutes to generate an adversarial example, which limits its wide use in the real world.

\subsection{Signal Processing-based Attacks}

Abdullah et al. \cite{abdullah2019practical} noticed that different audio samples may have the same feature vector when being transformed by acoustic feature extraction algorithms (e.g., MFCC) and nearly all SRSs appear to rely on several feature extraction algorithms. Based on this knowledge, they successfully attacked a commercial SVS and a commercial SIS in black-box settings. They first obtained desired audio (e.g., \emph{OK, Google} uttered by Alice). Then they designed four methods to obfuscate the desired audio as much as possible to generate obfuscated audio, i.e., adversarial examples. The four methods include Time Domain Inversion (TDI), Random Phase Generation (RPG), High Frequency Addition (HFA), and Time Scaling (TS). Although this work achieved efficient attacks since an adversarial example could be generated in a few seconds, adversarial examples it generated were noise in human perception which were easy to be noticed by human listeners.

Abdullah et al. \cite{abdullah2021hear} designed an attack method to circumvent surveillance in telephone networks. They assumed that SISs rely on the components of audio that are non-essential for human comprehension. In order to find out the non-essential components, they   first split audio into phonemes. Then, signal decomposition algorithms were used to decompose the phoneme into individual components and corresponding intensities. Since the low-intensity components are less perceptible to humans, the low-intensity components are likely what is looked for. Therefore, adversarial examples are generated by filtering out low-intensity components of every phoneme. They employed an untargeted attack in black-box settings to prove the effectiveness of the attack method they proposed. The results show that when only one phoneme in an utterance is perturbed, the ASR of the utterance can reach 10\%-20\% for most phonemes. An adversary can perturb multiple phonemes in an utterance to achieve a high ASR. They further demonstrated that the attack method is transferable in a cross-model setting. Although they achieved untargeted attacks and over-telephone-network attacks, most adversaries cannot obtain the right to monitor the telephone network. Targeted attacks and over-air attacks against intelligent voice assistants embedded in smart devices are the mainstream of adversarial attacks against SRSs now and even in the future.

While \cite{abdullah2021hear} successfully achieved untargeted attacks by removing low-intensity components of clean samples, Wang et al. \cite{wang2020inaudible} tried to achieve target attacks by adding additional components to clean samples. They also aimed to generate inaudible adversarial perturbations, instead of maintaining a slight noise to the clean sample. Inspired by previous work \cite{schonherr2018adversarial,qin2019imperceptible} on speech recognition systems, they   first obtained the desired audio (e.g., \emph{OK, Google}) as adversarial perturbations. Then, they leveraged frequency masking, which refers to the phenomenon that one faint but audible sound becomes inaudible in the presence of another louder audible sound, to hide adversarial perturbations in normal audio, such as birdsong and white noise. They successfully attacked a DNN-based SIS by inaudible adversarial audio in human perception. Unfortunately, they deployed experiments in white-box settings rather than more practical black-box settings. In addition, they did not explore the transferability of adversarial examples they generated and did not explore over-air attacks.

\subsection{Comparison and Discussion}

In Section 4, we comprehensively review adversarial attack methods in SRSs and compare them in Table 1. Based on Table 1, we conclude our review as below. 

Among all the works we reviewed in this section \cite{kreuk2018fooling,li2020adversarial,li2020universal,xie2020real,li2020advpulse,chen2021real,abdullah2019practical,abdullah2021hear,wang2020inaudible}, the methods based on optimization \cite{kreuk2018fooling,li2020adversarial,li2020universal,xie2020real,li2020advpulse,chen2021real} account for two-third, while the methods based on signal processing \cite{abdullah2019practical,abdullah2021hear,wang2020inaudible} account for one-third. The first type of method can be transferred from computer vision directly since they generate adversarial examples from digital vectors rather than raw audio or images. For example, Goodfellow et al. \cite{goodfellow2014explaining} proposed FGSM to deceive an image classification model, while Kreuk et al. \cite{kreuk2018fooling} and Li et al. \cite{li2020adversarial} also used FGSM to fool SRSs successfully.  The methods based on signal processing cannot be transferred in different domains directly since they leverage acoustic signal processing techniques to generate adversarial examples. For example, Abdullah et al. \cite{abdullah2021hear} generated adversarial examples by filtering out low-intensity components in raw audio. Obviously, this type of method cannot be used to deceive image processing tasks.

We observe that researchers not only focus on simple white-box attacks but also pay attention to black-box attacks. More than half of reviewed works \cite{kreuk2018fooling,li2020adversarial,chen2021real,abdullah2019practical,abdullah2021hear} explored the effectiveness of attack methods in black-box settings. In addition, all reviewed works except \cite{abdullah2021hear} achieved targeted attacks, which are more difficult than untargeted attacks.

Three of all reviewed works \cite{li2020universal,xie2020real,li2020advpulse} generated universal adversarial perturbations, which help adversaries to achieve practical real-time attacks. More than half of reviewed methods \cite{kreuk2018fooling,li2020adversarial,chen2021real,abdullah2019practical,abdullah2021hear} proved that adversarial examples they generated are transferable. Transferable adversarial examples can not only deceive the SRS that generates them, but also deceive other SRSs. In other words, transferable adversarial examples can deceive multiple SRSs, while non-transferable adversarial examples can only deceive the SRS that generates them. Therefore, transferable adversarial examples are more practical in the real world. In addition, four of all reviewed studies \cite{xie2020real,li2020advpulse,chen2021real,abdullah2019practical} considered over-air attacks and deployed experiments to explore over-air attacks. Among them, the effective attack distance of adversarial examples generated by \cite{abdullah2019practical} is only 0.3 meters, while the effective attack distance of adversarial examples generated by \cite{li2020advpulse} can reach 3 meters. More than half of reviewed attack methods \cite{li2020advpulse,chen2021real,abdullah2019practical,abdullah2021hear,wang2020inaudible} can successfully attack commercial SRSs.

Adversarial examples generated by the optimization-based attack methods \cite{kreuk2018fooling,li2020adversarial,li2020universal,xie2020real,li2020advpulse,chen2021real} sound clean to humans. Adversarial examples generated by the three attack methods based on signal processing \cite{abdullah2019practical,abdullah2021hear,wang2020inaudible} are clean, noisy, and inaudible to humans, respectively. Additionally, more than half of reviewed works \cite{kreuk2018fooling,li2020adversarial,chen2021real,abdullah2021hear,wang2020inaudible} deployed human perception experiments to explore the imperceptibility of adversarial perturbations. At the same time, four reviewed works \cite{li2020universal,xie2020real,li2020advpulse,chen2021real,wang2020inaudible} measured SNR of adversarial examples to quantitatively represent the relationship between audio and perturbations in adversarial examples.

Three of all reviewed works \cite{li2020universal,xie2020real,li2020advpulse} achieve real-time attacks since they generate universal adversarial perturbations. The optimization-based attack method proposed in \cite{chen2021real} takes several minutes to generate an adversarial example, while the signal processing-based attack methods proposed in \cite{abdullah2019practical,abdullah2021hear} only take several seconds. This indicates that the methods based on signal processing are more time-efficient than the optimization-based attack methods.

\newcommand{\tabincell}[2]{\begin{tabular}{@{}#1@{}}#2\end{tabular}}
\begin{sidewaystable}[!htbp] 
%\begin{table}[!htbp]
\caption{Comparison of Adversarial Attacks against SRSs}
\label{Comparison of Adversarial Attacks against SRSs}
\centering
\setlength{\tabcolsep}{1mm}{
\begin{tabular}{|c|c|c|c|c|c|c|c|c|c|c|c|c|c|c|c|c|c|c|}
\hline
\multirow{2}{*}{Ref}&
\multirow{2}{*}{Type}&
\multirow{2}{*}{Task}&
\multirow{2}{*}{AK-M}&
\multirow{2}{*}{AS}&
\multicolumn{5}{c|}{Practicability}&
\multicolumn{4}{c|}{Imperceptibility}&
\multicolumn{4}{c|}{Effectiveness}&
\multirow{2}{*}{Attack Method} \\
\cline{6-18}
 & & & & & Tr & Un & AM & Dis(m) & CS & ToA & PN & HP & SNR(dB) & GT & BA & AA & Met(\%) & \\
\hline
\multirow{2}{*}{\cite{kreuk2018fooling}}&
\multirow{2}{*}{O}&
\multirow{2}{*}{SVS}&
\multicolumn{1}{c|}{W}&
\multirow{2}{*}{Ta}&
\multicolumn{1}{c|}{?}&
\multirow{2}{*}{$\times$}&
\multirow{2}{*}{L}&
\multirow{2}{*}{-}&
\multirow{2}{*}{$\times$}&
\multirow{2}{*}{C}&
\multirow{2}{*}{$l_\infty$}&
\multirow{2}{*}{$\checkmark$}&
\multirow{2}{*}{?}&
\multirow{2}{*}{?}&
\multicolumn{1}{c|}{87.5/4.88}&
\multicolumn{1}{c|}{25.75/94.63}&
\multirow{2}{*}{RA/FPR}&
\multirow{2}{*}{FGSM} \\
\cline{4-4} \cline{6-6} \cline{16-17}
 & & & B & & $\checkmark$ & & & & & & & & & & 81.55/12 & 58.93/46 & & \\
\hline
\multirow{2}{*}{\cite{li2020adversarial}}&
\multirow{2}{*}{O}&
\multirow{2}{*}{SVS}&
\multicolumn{1}{c|}{W}&
\multirow{2}{*}{Ta}&
\multicolumn{1}{c|}{?}&
\multirow{2}{*}{$\times$}&
\multirow{2}{*}{L}&
\multirow{2}{*}{-}&
\multirow{2}{*}{$\times$}&
\multirow{2}{*}{C}&
\multirow{2}{*}{$l_\infty$}&
\multirow{2}{*}{$\checkmark$}&
\multirow{2}{*}{?}&
\multirow{2}{*}{?}&
\multicolumn{1}{c|}{7.2/7.2}&
\multicolumn{1}{c|}{96.87/97.64}&
\multicolumn{1}{c|}{FPR/EER}&
\multirow{2}{*}{FGSM} \\
\cline{4-4} \cline{6-6} \cline{16-18}
 & & & B & & $\checkmark$ & & & & & & & & & & 6.62 & 74.32 & EER & \\
\hline
\multirow{2}{*}{\cite{li2020universal}}&
\multirow{2}{*}{O}&
\multirow{2}{*}{SIS}&
\multirow{2}{*}{G}&
\multicolumn{1}{c|}{Uta}&
\multirow{2}{*}{?}&
\multirow{2}{*}{$\checkmark$}&
\multirow{2}{*}{L}&
\multirow{2}{*}{-}&
\multirow{2}{*}{$\times$}&
\multirow{2}{*}{C}&
\multirow{2}{*}{$l_2$}&
\multirow{2}{*}{?}&
\multicolumn{1}{c|}{44.13}&
\multirow{2}{*}{Real-time}&
\multirow{2}{*}{-}&
\multicolumn{1}{c|}{97.5}&
\multirow{2}{*}{ASR}&
\multirow{2}{*}{Generative Network} \\
\cline{5-5} \cline{14-14} \cline{17-17}
 & & & & Ta & & & & & & & & & 48.53 & & & 97.2 & & \\
\hline
\cite{xie2020real}&O&SIS&W&Ta&?&$\checkmark$&L,A&?&$\times$&C&?&?&9.42&real-time&-&90.19&ASR&Optimization+RIR\\
\hline
\cite{li2020advpulse}&O&SIS&W&Ta&?&$\checkmark$&L,A&1.6-3&$\checkmark$&C&$l_2$&?&8.3&real-time&-&96.9&ASR&AdvPulse\\
\hline
\cite{chen2021real}&O&\tabincell{c}{SIS,\\SVS}&B&Ta&$\checkmark$&$\times$&L,A&1&$\checkmark$&C&$l_\infty$&$\checkmark$&30.2&minutes&-&99&ASR&FakeBob\\
\hline
\cite{abdullah2019practical}&SP&\tabincell{c}{SIS,\\SVS}&B&Ta&$\checkmark$&$\times$&L,A&0.3&$\checkmark$&N&-&?&?&seconds&-&100&ASR&TDI,RPG,HFA,TS\\
\hline
\cite{abdullah2021hear}&SP&SIS&B&Uta&$\checkmark$&$\times$&L,TN&-&$\checkmark$&C&$L_2$&$\checkmark$&?&seconds&-&10-20&ASR& \tabincell{c}{Filtering Out Low-\\intensity Components}\\
\hline
\cite{wang2020inaudible}&SP&SIS&W&Ta&?&$\times$&L&-&$\checkmark$&In&$L_\infty$&$\checkmark$&34.111&?&-&93.8&ASR&Frequency Masking\\
\hline
\multicolumn{19}{|c|}{\tabincell{c}{AK-M: Adversary's Knowledge about Models; AS: Adversarial Specificity; Tr: Transferability; Un: Universality; AM: Attack Media; Dis: Distance;\\CS: Commercial SRSs; ToA: Types of Adversarial Audio; PN: Perturbation Norm; HP: Human Perception; SNR: Signal-to-noise Ratio;\\GT: Generation Time; BA: Before Attack; AA: After Attack; Met: Metrics; O: Optimization; SP: Signal Processing; W: White-box; G: Grey-box;\\B: Black-box; Ta: Targeted; Uta: Untargeted; L: Line; A: Air; TN: Telephone Network; C: Clean; N: Noisy; In: Inaudible; $\checkmark$: satisfied;\\$\times$: \textup{not satisfied}; ?: \textup{not discussed}; -: \textup{not available}.}} \\
\hline
\end{tabular}}
%\end{table}
\end{sidewaystable}

\section{Defenses against SRSs}

In this section, we first propose a taxonomy of existing defenses against SRSs. After that, we review and evaluate some defense methods against adversarial attacks in SRSs (as shown in Table 2) by applying the criteria proposed in Section 3.2.

\subsection{Taxonomy of Defenses against SRSs}

There are two types of defense methods against adversarial attacks: 1) proactive defenses, 2) passive defenses. Proactive defense methods employ adversarial data augmentation to retrain original models such that they can be robust to adversarial examples. Passive defense methods defend against adversarial attacks by adding new components rather than modifying original models. According to the function of new components, passive defense methods can be divided into detection methods and purification methods. When an adversarial example is identified, a detection method aims to refuse it to enter systems, while a purification method aims to feed it to systems after removing adversarial perturbations.

\subsection{Proactive Defenses}

Wang et al. \cite{wang2019adversarial} proposed adversarial regularization based on adversarial examples to defend against adversarial attacks. Adversarial regularization aims to seek the worst sample around an input sample and then use the worst sample to optimize an SRS. They used adversarial examples generated by FGSM and virtual adversarial training based on local distributional smoothness (LDS) to attack a DNN-based SVS. It is worth noting that virtual adversarial training can calculate adversarial perturbations for unlabeled samples. To the best of our knowledge, this is the first work to apply virtual adversarial training into SVSs. After that, they leveraged FGSM and LDS to regularize the SVS, respectively. They showed that adversarial regularization is a medium-general defense method through several experiments. However, the performance of adversarial regularization is not good enough, since the EER of the regularized SVS only decreased slightly. In addition, this work did not consider adaptive attacks and did not mention defense time.

Many previous works have explored defense methods to resist spoofing attacks, including synthesis, convert and replay attacks, for SVSs. However, spoofing countermeasure models are still vulnerable to adversarial attacks. To address this issue, Wu et al. \cite{wu2020defense} proposed two defense methods, one is proactive, i.e., adversarial training, and the other is passive, i.e., spatial smoothing, to improve the robustness of SVS spoofing countermeasure models. We will introduce spatial smoothing in Subsection 5.3. Adversarial training refers to a defense method that uses adversarial data augmentation to retrain the model to enhance the robustness of the model. They retrained a DNN-based SVS by adversarial data augmentation which was generated by Projected Gradient Descent (PGD) method. They proved that the performance of adversarial training is better than spatial smoothing through several experiments, which is because adversarial training is model-specific. However, adversarial training is low-generality. In addition, defense time and adaptive attacks were not mentioned in the paper.

\subsection{Passive Defenses}
Passive defense methods utilize new components to detect or purify adversarial examples. In this subsection, we first review detection methods, followed by a review of purification methods.

\subsubsection{Detection Methods}

Although adversarial training is effective, it is difficult to obtain adversarial data augmentation since we need to label every adversarial example. Inspired by \cite{gong2017adversarial,samizade2020adversarial}, Li et al. \cite{li2020investigating} made the first attempt to defend SVSs against adversarial attacks with a separate detection network which is a VGG-like network structure. The separate detection network not only avoids retraining well-developed SVSs but also can combine with countermeasures against spoofing attacks to obtain a powerful defense method. They first respectively adjusted parameters of two separate detection networks using adversarial examples generated by BIM and Jacobian-based Saliency Map Attack (JSMA) to obtain two different separate detection networks, i.e., a BIM-based separate detection network and a JSMA-based separate detection network. Then, they proved that the BIM-based separate detection network can not only detect adversarial examples generated by BIM but also detect adversarial examples generated by JSMA to a certain extent. Similarly, the JSMA-based separate detection network can also detect adversarial examples generated by BIM to a certain extent. In other words, the defense method they proposed, i.e. the separate detection network, is effective and medium-general. However, they did not consider adaptive attacks, which are more challenging, and also did not mention defense time.

The above defense methods \cite{wu2020defense,li2020investigating,wang2019adversarial} require knowledge of the attack methods used by adversaries. However, it is impractical for SRSs designers to know which attack methods will be implemented by adversaries in advance. Therefore, Wu et al. \cite{wu2021voting} proposed a highly general defense method called voting for the right answer. It means that whether an input utterance is accepted by the SVS is determined by the similarity between the input utterance and the enrollment utterance and the similarities between the enrollment utterance and neighbors of the input utterance which are some samples randomly selected around the input utterance. As its name suggests, this method means that the input utterance and its neighbors are voting on whether to accept the input. They used adversarial examples generated by BIM to attack a DNN-based SVS in white-box settings and considered both adaptive attacks and non-adaptive attacks. Although the proposed defense method is simple and effective, there are still two issues: 1) the performance of the defense method is related to some parameters that are difficult to select; 2) during defense, the SVS needs to run many times for every sample since we need to calculate the similarities between its neighbors and the enrollment utterance, which is time-consuming; 3) they did not discuss defense time quantitatively.

Wu et al. \cite{wu2021spotting} also proposed a highly general defense method. They leveraged Parallel WaveGan, a neural vocoder, to re-synthesize the input utterance, and then used the difference between the SVS scores, i.e., the similarity with the enrollment utterance, for the input and re-synthesized utterance to determine whether the input utterance is an adversarial example. Since neural vocoders can purify adversarial perturbations, the large difference in SVS scores indicates the input utterance is an adversarial example. This is the first work to adopt neural vocoders as shields to detect adversarial examples for SVSs, and it showed neural vocoders are effective to detect adversarial examples by several experiments. Meanwhile, this work also clarified by experiments that the defense method slightly affected clean samples. However, it did not analyze adaptive attacks and defense time.

\subsubsection{Purification Methods}

Inspired by \cite{dong2020secure}, Wu et al. \cite{wu2020defense} proposed a highly general defense method based on spatial smoothing. The reason is that implementing smoothing does not need extra training efforts. In image processing, spatial smoothing uses nearby pixels to smooth the central pixel. According to different weighting mechanisms of nearby pixels, spatial smoothing can be divided into many categories, such as median filter, mean filter, and Gaussian filter. The authors leveraged these filters for SVSs, and employed experiments to prove the effectiveness of spatial smoothing. They further explored the combination of spatial smoothing and adversarial training, which achieves better performance. However, they ignored adaptive attacks and did not analyze defense time.

Zhang et al. \cite{zhang2020adversarial} designed an adversarial separation network (AS-Net) to defend against adversarial attacks in SRSs. AS-Net aims to eliminate adversarial perturbations and restore natural clean utterances. Two optimized components, including compression structure and speaker quality loss, are introduced. The former is responsible for reconstructing adversarial perturbations, and the latter supervises whether the restored utterances generated by AS-Net are correctly labeled by the target SRS. They deployed a lot of experiments to defend against FGSM, PGD, Decoupled Direction and Norm (DDN) and Momentum attack (MT), which show the effectiveness and medium-generality of AS-Net in DNN-based SRSs. In addition, they compared the performance of different countermeasures, including adversarial training, feature-squeezing, and AS-Net. The results show that AS-Net significantly outperformed other countermeasures. However, they did not consider defense time and ignored adaptive attacks.

Wu et al. \cite{wu2021adversarial} proposed a highly general defense method based on cascaded self-supervised learning models, which possesses the ability to mitigate superficial perturbations in the input utterance after pretraining. Transformer encoder representations from alteration (TERA) as an advanced self-supervised learning model was used to construct the defense method. They generated adversarial examples by BIM to attack a DNN-based SVS. It is worth noting that they considered both adaptive attacks and non-adaptive attacks. The results show the defense method is effective on both of them to some extent and it is more difficult to defend against adaptive attacks. However, the experimental results also show that the defense method they proposed has a negative impact on clean samples. In addition, defense time was neglected.

\subsection{Comparison and Discussion}

In Section 5, we comprehensively review the existing works about defense methods \cite{wang2019adversarial,wu2020defense,li2020investigating,wu2021voting,wu2021spotting,zhang2020adversarial,wu2021adversarial} for adversarial attacks in SRSs. Meanwhile, we compare all the defense methods reviewed in this section in Table 2. Based on Table 2, we summarize our review as below.

Among all the studies reviewed in this section, passive defense methods \cite{wu2020defense,li2020investigating,wu2021voting,wu2021spotting,zhang2020adversarial,wu2021adversarial} are distinctly overwhelming with three-quarters of all reviewed papers. Passive defense methods are so popular since they can be deployed in any SRSs to defend against adversarial attacks. However, proactive defense methods \cite{wang2019adversarial,wu2020defense} cannot be transferred between different SRSs since they are model-specific, which makes them receive less attention than passive defense methods.

We observe that all reviewed works deployed experiments to defend against non-adaptive attacks, while only two works \cite{wu2021voting,wu2021adversarial} try to defend against adaptive attacks that are more threatening than non-adaptive attacks.

Half of reviewed defense methods \cite{wu2020defense,wu2021voting,wu2021spotting,wu2021adversarial} are highly general. Highly general defense methods are favored by researchers since they can defend against any attack methods theoretically. In addition, all reviewed works considered and deployed over-line attacks. Over-line ensures lossless transmission of adversarial examples. Therefore, defense methods that can defend against over-line attacks can also defend against over-air and over-telephone-network attacks.

All reviewed works employ experiments to show the effectiveness of defense against conventional attack methods, such as FGSM, BIM, and PGD. However, they do not defend against some advanced attacks, such as AdvPluse \cite{li2020advpulse} and FakeBob \cite{chen2021real}. At last, defense time, an important evaluation criterion of practicability, is missed in discussion in all reviewed works..

%\newcommand{\tabincell}[2]{\begin{tabular}{c}{@{}#1@{}}#2\end{tabular}}
%\begin{sidewaystable}[!htbp] 
\begin{table}[!htbp]
\caption{Comparison of Defense Methods against SRSs}
\label{Comparison of Defense Methods against SRSs}
\centering
\setlength{\tabcolsep}{1mm}{
\begin{tabular}{|c|c|c|c|c|c|c|c|c|c|c|c|c|}
\hline
\multirow{2}{*}{Ref}&
\multirow{2}{*}{Type}&
\multirow{2}{*}{Task}&
\multirow{2}{*}{\tabincell{c}{AK-\\DM}}&
\multicolumn{3}{c|}{Practicability}&
\multicolumn{5}{c|}{Effectiveness}&
\multirow{2}{*}{Defense Method} \\
\cline{5-12}
 & & & & Ge & DM & DfA & DT & Ori & AA & AD & Met(\%) &  \\
\hline
\cite{wang2019adversarial} & Pro & SVS & Nad & Me & L & FGSM,LDS & ? & 4.87 & 11.89 & 8.31 & EER & Adversarial Regularization\\
\hline
\cite{wu2020defense}& Pro & SVS & Nad & Lo & L & PGD & ? & 99.99 & 37.06 & 98.60 & DA & Adversarial Training\\
\hline
\cite{li2020investigating} & P-D & SVS & Nad & Me & L & BIM,JSMA & ? & \tabincell{c}{5.97/\\-} & \tabincell{c}{39.87/\\-} & \tabincell{c}{0.18/\\99.83} & \tabincell{c}{EER/\\DA} & \tabincell{c}{Separate Detection\\Network}\\
\hline
\multirow{3}{*}{\cite{wu2021voting}}&
\multirow{3}{*}{P-D}&
\multirow{3}{*}{SVS}&
\multicolumn{1}{c|}{Nad}&
\multirow{3}{*}{Hi}&
\multirow{3}{*}{L}&
\multirow{3}{*}{BIM}&
\multirow{3}{*}{?}&
\multirow{3}{*}{\tabincell{c}{2.24/\\2.56}}&
\multirow{3}{*}{\tabincell{c}{71.83/\\74.92}}&
\multicolumn{1}{c|}{\tabincell{c}{10.66/\\24.68}}&
\multirow{3}{*}{\tabincell{c}{FPR/\\FNR}}&
\multirow{3}{*}{\tabincell{c}{Voting for the\\Right Answer}}\\
\cline{4-4} \cline{11-11} 
 & & & Ad &  & & & & &  & \tabincell{c}{13.29/\\27.75} & &\\
\hline
\cite{wu2021spotting}& P-D& SVS  & Nad & Hi & L & BIM & ? & \tabincell{c}{2.88/\\-} & \tabincell{c}{99.33/\\-} & \tabincell{c}{-/\\99.76} & \tabincell{c}{EER/\\DA} & Neural Vocoders\\
\hline
\cite{wu2020defense} & P-P& SVS & Nad & Hi & L & PGD & ? & 99.99 & 48.32 & 93.95 & DA & Spatial Smoothing\\
\hline
\cite{zhang2020adversarial} & P-P & \tabincell{c}{SVS,\\SIS}& Nad & Me & L & \tabincell{c}{FGSM,MT,\\PGD,DDN} & ? & 0.89 & 13.81 & 3.62 & EER & \tabincell{c}{Adversarial Separate\\Network}\\
\hline
\multirow{2}{*}{\cite{wu2021adversarial}}&
\multirow{2}{*}{P-P}&
\multirow{2}{*}{SVS}&
\multicolumn{1}{c|}{Nad}&
\multirow{2}{*}{Hi}&
\multirow{2}{*}{L}&
\multirow{2}{*}{BIM}&
\multirow{2}{*}{?}&
\multirow{2}{*}{8.87}&
\multirow{2}{*}{66.02}&
\multicolumn{1}{c|}{22.94}&
\multirow{2}{*}{EER}&
\multirow{2}{*}{\tabincell{c}{Cascaded Self-supervised\\Learning Models}}\\
\cline{4-4} \cline{11-11} 
 & & & Ad &  & & & & &  & 40.69 & &\\
\hline
\multicolumn{13}{|c|}{\tabincell{c}{AK-DM: Adversary's Knowledge about Defense Methods; Ge: Generality; DM: Defense Media;\\DfA: Defendable Attacks; DT: Defense Time; Ori: Original; AA: After Attack; AD: After Defense;\\Met: Metrics; Pro: Proactive; P-D: Passive-Detection; P-P: Passive-Purification; Ad: Adaptive;\\Nad: Non-adaptive; Hi: High; Me: Medium; Lo: Low; L: Line; ?: not discussed; -: not available.} } \\
\hline
\end{tabular}}
\end{table}
%\end{sidewaystable}

\section{Open Issues and Future Directions}

\subsection{Open Issues}

By reviewing and comparing the above literature with our proposed criteria, we figure out several open issues for adversarial attacks and defenses in SRSs.

First, it is difficult to enhance the robustness of SRSs. Adversarial training, which utilizes adversarial data augmentation to retrain SRSs, is an effective way to enhance the robustness of SRSs. However, obtaining adversarial data augmentation, i.e., adversarial example and its true label pairs, is time-consuming since we need to manually label each adversarial example. Therefore, how to enhance the robustness of SRSs efficiently is still an open and tough issue.

Second, it is not convenient to directly compare the performance of attack methods or defense methods proposed in different works. This is caused by the differences in experimental settings and evaluation metrics applied in different works. For example, the defense methods proposed in \cite{wu2021voting} and \cite{wu2021spotting} were used to defend against adversarial examples generated by BIM. However, we cannot directly compare the performance of them since \cite{wu2021voting} and \cite{wu2021spotting} used different evaluation metrics, i.e., \cite{wu2021voting} used FPR and FAR, and \cite{wu2021spotting} used EER and DA. Similarly, we cannot compare the performance of defense methods proposed in \cite{zhang2020adversarial} and \cite{wu2021adversarial} since they were used to defend against adversarial examples generated by different attack methods. All in all, the differences in experimental settings and evaluation metrics among different works hinder researchers from comparing the performance of attacks methods and defense methods in a direct way. Uniform evaluation metrics or criteria should be defined and adopted.

Third, signal processing-based attacks receive little attention. On one hand, although signal processing-based attacks are more efficient than optimization-based attacks as discussed in Subsection 5.4, to the best of our knowledge, only three articles \cite{abdullah2019practical,abdullah2021hear,wang2020inaudible} raise signal processing-based attacks in SRSs, which are far less than optimization-based attacks. On the other hand, all defense methods are proposed to defend against adversarial examples generated by optimization-based attacks, such as FGSM and BIM, while ignoring signal processing-based attacks. We cannot judge if current defense methods can defend against signal processing-based attacks. In short, signal processing-based attacks have not been paid sufficient attention in the current literature.

Fourth, the literature still lacks research on poisoning attacks against SRSs. The poisoning attacks refer to adding malicious data into training data, resulting in a biased model. Previous work has shown that poisoning attacks can seriously threaten the security and privacy of ML models in the image field \cite{aghakhani2021bullseye, shafahi2018poison,zhu2019transferable}. However, as the structure of SRSs is more complex than image processing systems, little work pays attention to poisoning attacks against SRSs.

\subsection{Future Directions}

We suggest several future research directions motivated by the above open issues as below.

Firstly, applying virtual adversarial training \cite{miyato2018virtual} to enhance the robustness of SRSs is worthy of deep-insight research. Virtual adversarial training can relieve the pressure of labeling adversarial examples since it retrains SRSs in semi-supervised settings. In addition, virtual adversarial training has low computational costs and a small number of hyperparameters. Therefore, virtual adversarial training in SRSs may be an interesting attempt to enhance the robustness of SRSs. 

Secondly, unified adversarial attack and defense evaluation frameworks should be established. Specifically, the attack evaluation framework should clarify target SRSs, which include both open source systems and commercial systems, and attack evaluation metrics as shown in Subsection 3.1.3. The defense evaluation framework should include defendable attacks, which are used to generate adversarial examples for evaluating the performance of defense methods, and defense evaluation metrics as shown in Subsection 3.2.3. To comprehensively evaluate the performance of defense methods, both optimization-based attacks and signal processing-based attacks should be included into defendable attacks. In short, establishing unified frameworks for both attack and defense is an effective way to help researchers compare the performance of different methods of adversarial attack and defense to stimulate mutual development.

Thirdly, the research on the security of preprocessing module and feature extraction module should be strengthened. As shown in Fig.1, the SRS includes three modules, i.e., preprocessing module, feature extraction module and model inference module. The structure of SRSs is more complex than image recognition systems due to the addition of preprocessing and feature extraction. Each module introduces a surface of attacks, causing exploitable vulnerabilities from the perspective of an adversary. Researchers in the audio field proposed some signal processing-based attacks that utilize vulnerabilities of the preprocessing module or the feature extraction module, such as those mentioned in \cite{abdullah2019practical,abdullah2021hear,wang2020inaudible} for SRSs and in \cite{vaidya2015cocaine,carlini2016hidden} for speech recognition systems. However, there are still many unknown vulnerabilities. Therefore, we recommend strengthening the vulnerability mining of preprocessing module and feature extraction module. Correspondingly, defense methods should also be studied to enhance the robustness of these two modules. We believe that such an arms race will promote the security of SRSs.

Finally, we suggest studying poisoning attacks against SRSs and corresponding defense methods. On one hand, state-of-the-art SRSs require a huge amount of training data and it is common to collect these data from potentially untrustworthy sources (e.g., edge devices). Therefore, it is easy to poison the training dataset of SRSs. On the other hand, federated learning is a popular method to train ML models in a somehow privacy-preserving way, including ML models used in SRSs. However, it is difficult to guarantee that each party participating in federated learning is honest and trustworthy. A malicious party may deliberately use poisoned data for model training resulting in security and privacy threats. Therefore, it is interesting and promising to research poisoning attacks against SRSs and corresponding defense methods, especially in the context of federated learning, as well as other learning models.

\section{Conclusion}

In this paper, we overviewed the adversarial attacks and attack countermeasures in SRSs. We proposed two sets of criteria to evaluate the performance of adversarial attacks and defense methods. Based on our proposed taxonomies of existing adversarial attacks and defense methods, we reviewed existing adversarial attacks and defense methods by employing our proposed criteria, respectively. Through thorough review and analysis, we figured out several open research issues and highlighted future research directions to motivate the research of SRSs security.

\section{Acknowledge}

This work is supported in part by the National Natural Science Foundation of China under Grant 62072351; in part by the Academy of Finland under Grant 308087, Grant 335262, Grant 345072 and Grant 350464; in part by the open research project of ZheJiang Lab under grant 2021PD0AB01; in part by the Shaanxi Innovation Team Project under Grant 2018TD-007; and in part by the 111 Project under Grant B16037.

\bibliography{mybibfile}

\end{document}